\documentclass[aps,prl,superscriptaddress,showpacs,twocolumn]{revtex4-1}
 \usepackage{amsmath}  % ,twocolumn
 \usepackage{makeidx}
 \usepackage{graphicx}
 \usepackage{amsfonts}
 \usepackage[ansinew]{inputenc}
 \usepackage[usenames,dvipsnames]{pstricks}
 \usepackage{subfigure}
 \usepackage{footnote}
 \usepackage{CJK}
 \usepackage{lipsum}
 \usepackage{multirow}
 \usepackage{mathrsfs}
 \allowdisplaybreaks[1]

              \makeindex

\begin{document}

\begin{CJK*}{GB}{}

\title{$\alpha$-Clustering in Nuclei and Its Impact on Nuclear Symmetry Energy}

\author{Shuo Yang }
\affiliation{School of Physics, Nanjing University, Nanjing 210093, China}
%({\CJKfamily{gbsn}��˶})

\author{Ruijia Li }
\affiliation{School of Physics, Nanjing University, Nanjing 210093, China}

\author{Chang Xu }
\email{cxu@nju.edu.cn}
\affiliation{School of Physics, Nanjing University, Nanjing 210093, China}
\affiliation{Institute for Nonperturbative Physics, Nanjing University, Nanjing 210093, China}
%({\CJKfamily{gbsn}����})
%\date{\today}

\begin{abstract}
Nuclear symmetry energy is a fundamental quantity currently under intense investigation in both nuclear physics and astrophysics. The {\it softness} or {\it stiffness} of symmetry energy is still under debate and the extraction of symmetry energy from neutron skin thickness $R_{\rm skin}$ remains a challenge. Parity-violating measurements PREX and CREX provide important opportunities for constraining $R_{\rm skin}$ in $^{208}$Pb and $^{48}$Ca. We investigate the occurrence of $\alpha$-cluster at the surface of nuclei and its impact on the extraction of symmetry energy from $R_{\rm skin}$. Our result indicates that the $\alpha$-clustering probability in $^{208}$Pb is small and the extracted density slope of symmetry energy $L$ is almost unaffected. In contrast, the $\alpha$-clustering probability in $^{48}$Ca is sizeable and the corresponding correction to $L$ should be taken into account. This correction progressively increases with the $\alpha$-clustering probability, leading to a modification of the $L$-$R_{\rm skin}$ correlation, a fact may have important implications in constraining nuclear symmetry energy.
\end{abstract}


\maketitle

\end{CJK*}

{\it Introduction.}- The formation of compact clusters ({\it e.g.} $\alpha$-clusters) is an interesting feature of nuclear quantum many-body system and plays an essential role in many important problems of astrophysics. The phenomena of $\alpha$-clustering are abundant in excited states of light nuclei close to decay threshold \cite{Freer2018}. One of the famous instances is the 3$\alpha$-structure Hoyle state in $^{12}$C, which unlocks the puzzle in the production of heavy elements inside stars \cite{THSR,THSR2017}. In contrast to light nuclei, the $\alpha$-clustering problem in heavy nuclei is still not fully solved and the microscopic treatment of cluster dynamics beyond mean-field theory is a great challenge \cite{review1,review2,review3,Denisov,Mohr,Royer,Poenaru}.

Recently, the PREX collaboration reported the measurement of parity-violating asymmetry $A_{PV}$ and deduced a rather large neutron skin thickness $R_{\rm skin}$ in $^{208}$Pb (PREX-2) \cite{PREX2}, Thick skin in $^{208}$Pb suggests a very stiff symmetry energy in contrast to the previous constraints obtained from many other observations \cite{Horowitz2012,Piekarewicz2021,Reed2021,Reinhard2021,Thomas2022,Yue2022}.
Very recently, the CREX collaboration has successfully conducted the parity-violating experiment in $^{48}$Ca and deduced a thin $R_{\rm skin}$ \cite{CREX}, suggesting a soft symmetry energy. Much effort has been expended on attempting to reconcile these seemingly contradictory results. Special attention has been devoted to the problem of $\alpha$-clustering at the surface of nuclei that is expected to affect the density slope of symmetry energy $L$ \cite{Essick2021}.

$L$ is critical for understanding not only the structure of rare isotopes and the reaction mechanism of heavy-ion collisions, but also the structure and the composition of neutron stars \cite{Typel2001,Agrawal2012,Piekarewicz2012,BALi2008,Tsang2009,Brown2000,Horowitz2001,Rutel2005,Vinas2014,Maza2015}. By using the Hugenholtz-Van Hove (HVH) theorem, $L$ can be decomposed in a unique way into kinetic energy, isoscalar potential and isovector potential contributions \cite{symmetry,symmetry2014}. As a fundamental relation for interacting self-bound infinite Fermi system, this theorem does not depend upon the precise nature of the interaction. While the kinetic energy and isoscalar potential contributions are relatively well constrained, the isovector potential contribution still has significant uncertainties. Owing to the existence of isovector potential, more neutrons are pushed from the inner region of finite nuclei outwards to the surface region, and thus contribute to $R_{\rm skin}$. In this sense, $L$ is related intrinsically to $R_{\rm skin}$.

The correlation between $R_{\rm skin}$ and $L$ could be modified by the occurrence of $\alpha$-clusters \cite{Typel}. This is because the $\alpha$-clusters may appear in the low density region, {\it i.e.} the surface of finite nucleus,  and its impact progressively increases with the $\alpha$-clustering probability \cite{quartet2014,Xu2016,Xu2017,qwfa2020,qwfa2021,48Ti,Science}. Inside the core, $\alpha$-clustering is suppressed and its four nucleons are considered to move almost independently in a shell-model mean-field potential. The single-particle states are populated up to the Fermi energies of the neutrons or protons and pairing correlation exists among the single-particle orbits. Pairing remains at high densities but $\alpha$-cluster dissolves and its four nucleons (2n+2p) turn into the single-particle motions forming the continuum of scattering states.

In this Letter, we address the question of whether $\alpha$-clustering at the surface of $^{208}$Pb and $^{48}$Ca has certain impact on  $R_{\rm skin}$ and $L$. We use the quartetting wave function approach (QWFA) to do so because it treats correctly both the intrinsic motion between four nucleons in the $\alpha$-cluster and the relative motion of the $\alpha$-cluster versus the core \cite{Xu2016,Xu2017,qwfa2020,qwfa2021}. Strong closed shell structure effects and complex derivative terms of the intrinsic wave function are properly taken into account in QWFA. The key quantity for clustering modification on the $R_{\rm skin}$-$L$ correlation is the $\alpha$-cluster formation probability, which is quantitatively obtained by solving the coupled equations of a first-principle approach to nuclear many-body systems without adjusting any parameter.

{\it Intrinsic energy of $\alpha$-cluster embedded in nuclear medium.}- Firstly, we simulate the $\alpha$-cluster formation at the surface of core nucleus by considering low-density nuclear medium, in which the $\alpha$-like four-nucleon correlations are described by the in-medium Schr\"{o}dinger equation. The corresponding wave function of four nucleons is decomposed into a center of mass (c.o.m.) motion part $\Psi^{\text{com}}$ and an intrinsic motion part $\varphi^{\text{intr}}$, which are coupled together with complex gradient terms. Such gradient terms are difficult to handle, but vanish in the case of homogeneous nuclear medium. With the Jacobian momenta ${\bf p}_{1}={\bf P}/4+{\bf k}/2+{\bf k}_{12}, {\bf p}_{2}={\bf P}/4+{\bf k}/2-{\bf k}_{12}, {\bf p}_{3}={\bf P}/4-{\bf k}/2+{\bf k}_{34},{\bf p}_{4}={\bf P}/4-{\bf k}/2-{\bf k_{34}}$, the in-medium wave equation for the intrinsic motion is reduced to \cite{quartet2014}
\begin{widetext}
\begin{eqnarray}
\frac{\hbar^2}{2m}[k^2\!+\!2 k_{12}^2\!+\!2 k_{34}^2]\varphi^{\text{intr}}({\bf k},{\bf k}_{12},{\bf k}_{34})\!+\!\int\frac{d^3 k'}{(2\pi)^3}\frac{d^3 k'_{12}}{(2\pi)^3}\frac{d^3 k'_{34}}{(2\pi)^3} {V}_4 \varphi^{\text{intr}}({\bf k}',\!{\bf k}'_{12},\!{\bf k}'_{34})=({W}^{\text{ext}}\!+\!{W}^{\text{intr}})\varphi^{\text{intr}}({\bf k},\!{\bf k}_{12},\!{\bf k}_{34}),
\end{eqnarray}
\end{widetext}
where the centroid of $\alpha$-cluster is considered to be at rest ({\bf P}=0). ${V}_4$ is the effective in-medium interaction that contains the external mean field ${V}_4^{\rm ext}$ as well as the intrinsic NN interaction modified by the Pauli blocking ${V}_4^{\rm intr}
=\Theta(p_1-k_F)\Theta(p_2- k_F) V_{NN}({\bf p}_1,{\bf p}_2;{\bf p}'_1,
{\bf p}'_2)\delta({\bf p}_3-{\bf p}'_3)\delta({\bf p}_4-{\bf p}'_4)
+ {\text{5 permutations}}$. The NN interaction is defined as a Gaussian form factor $V_{NN}({\bf p}_1, {\bf p}_2;{\bf p}'_1,{\bf p}'_2)=\lambda {\rm e}^{-\frac{({\bf p}_1-{\bf p}_2)^2}{4 \gamma^2}} {\rm e}^{-\frac{({\bf p}'_1-{\bf p}'_2)^2}{4
\gamma^2}}\delta({\bf p}_1+{\bf p}_2-{\bf p}'_1-{\bf p}'_2)$ where the potential parameters $\lambda$ = 1449.6 MeV ${\rm fm}^3$ and $\gamma$ = 1.152 ${\rm fm}^{-1}$ \cite{quartet2014}.
\begin{figure}[htb]
\begin{minipage}{1\linewidth}
\centerline{\includegraphics[width=0.95\textwidth]{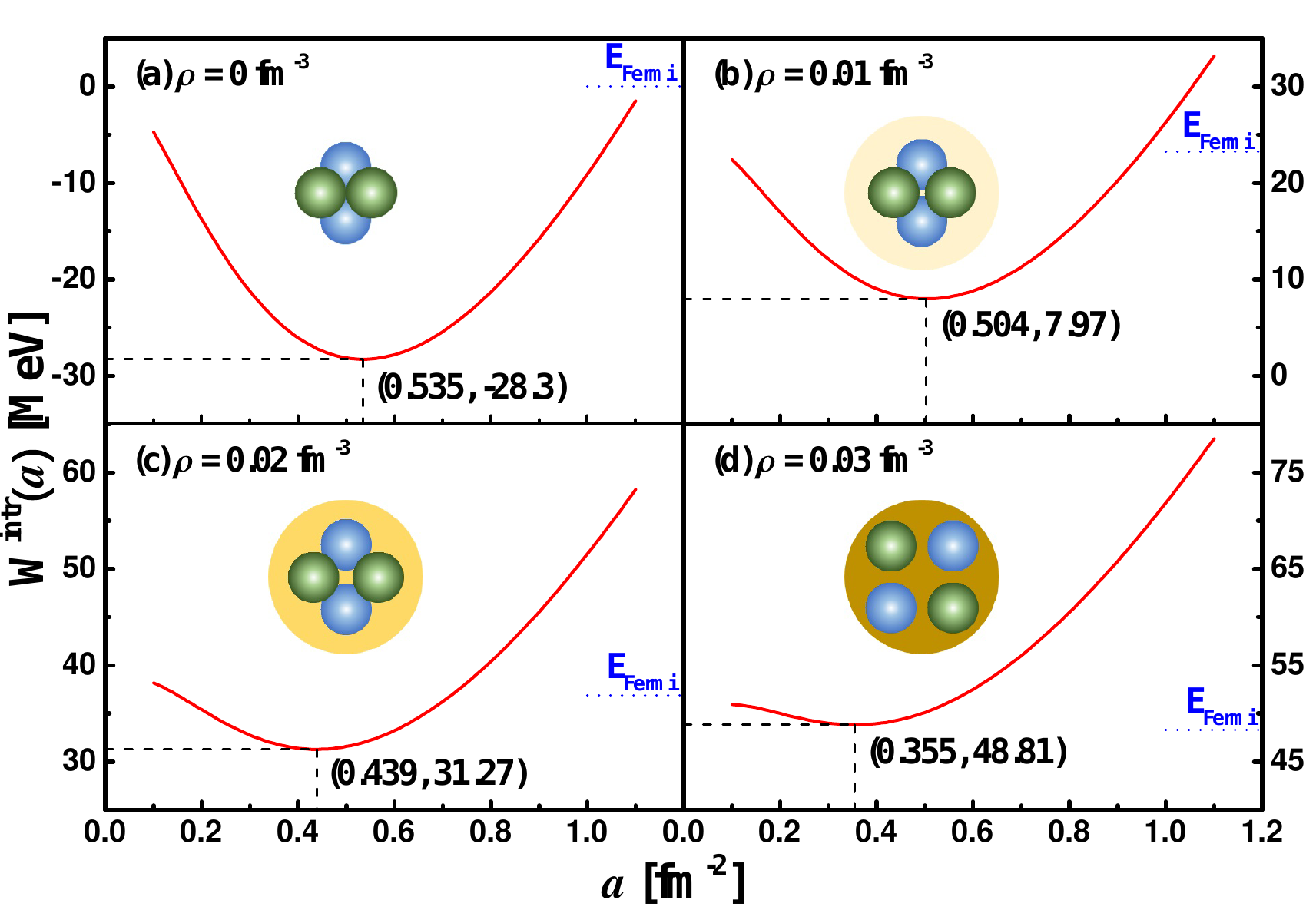}}
\end{minipage}
\caption{The variation of intrinsic energies of an $\alpha$-cluster in free space (a) and in homogeneous nuclear matters (b)-(d). A critical transition occurs at $\rho_c=0.03$ fm$^{-3}$ where the $\alpha$-cluster dissolves and its four nucleons become uncorrelated (d).}\label{Wa}
\end{figure}

The minimum of the intrinsic energy ${W}^{\text{intr}}$ has to be found for each density $\rho$ with the Fermi-blocked Gaussian ansatz $\varphi^{\rm intr}({\bf p}_1,{\bf p}_2,{\bf p}_3, {\bf p}_4)=\frac{1}{\sqrt{N}}
\varphi_{\tau_1}({\bf p}_{1})\varphi_{\tau_1}({\bf p}_{2})\varphi_{\tau_1}({\bf p}_{3})\varphi_{\tau_1}({\bf p}_{4}) \delta({\bf p}_1+{\bf p}_2+{\bf p}_3+{\bf p}_4)$ where $N$ is the normalization factor. The single nucleon wave function $\varphi_{\tau}({\bf p})$ is given by ${\rm e}^{-\frac{{\bf p}^2}{2a}} \Theta\left[p-k_F\right]$ with the single variational parameter $a$. The minimum energy of a free $\alpha$-cluster is ${W}^{\text{intr}}=-28.3$ MeV at $a=0.535$ fm$^{-2}$ (see Fig.\ref{Wa}(a)). The intrinsic energy ${W}^{\text{intr}}$ is shifted at finite density of the surrounding nuclear matter owing to the Pauli blocking. The bound state disappears and four nucleons become uncorrelated at the critical density $\rho_c=0.03$ fm$^{-3}$ (see Fig.\ref{Wa}(d)). Note that the matter density distribution of the $\alpha$-cluster at the surface of finite nucleus depends also on the surrounding density $\rho_{\alpha}(r,\rho)=4(\frac{4a(\rho)}{3\pi})^{3/2}{\rm e}^{-\frac{4a(\rho)}{3}r^2}$ by treating correctly both the energy shift and Pauli blocking effect.
\begin{figure}[htb]
\begin{minipage}{1\linewidth}
\centerline{\includegraphics[width=1.0\textwidth]{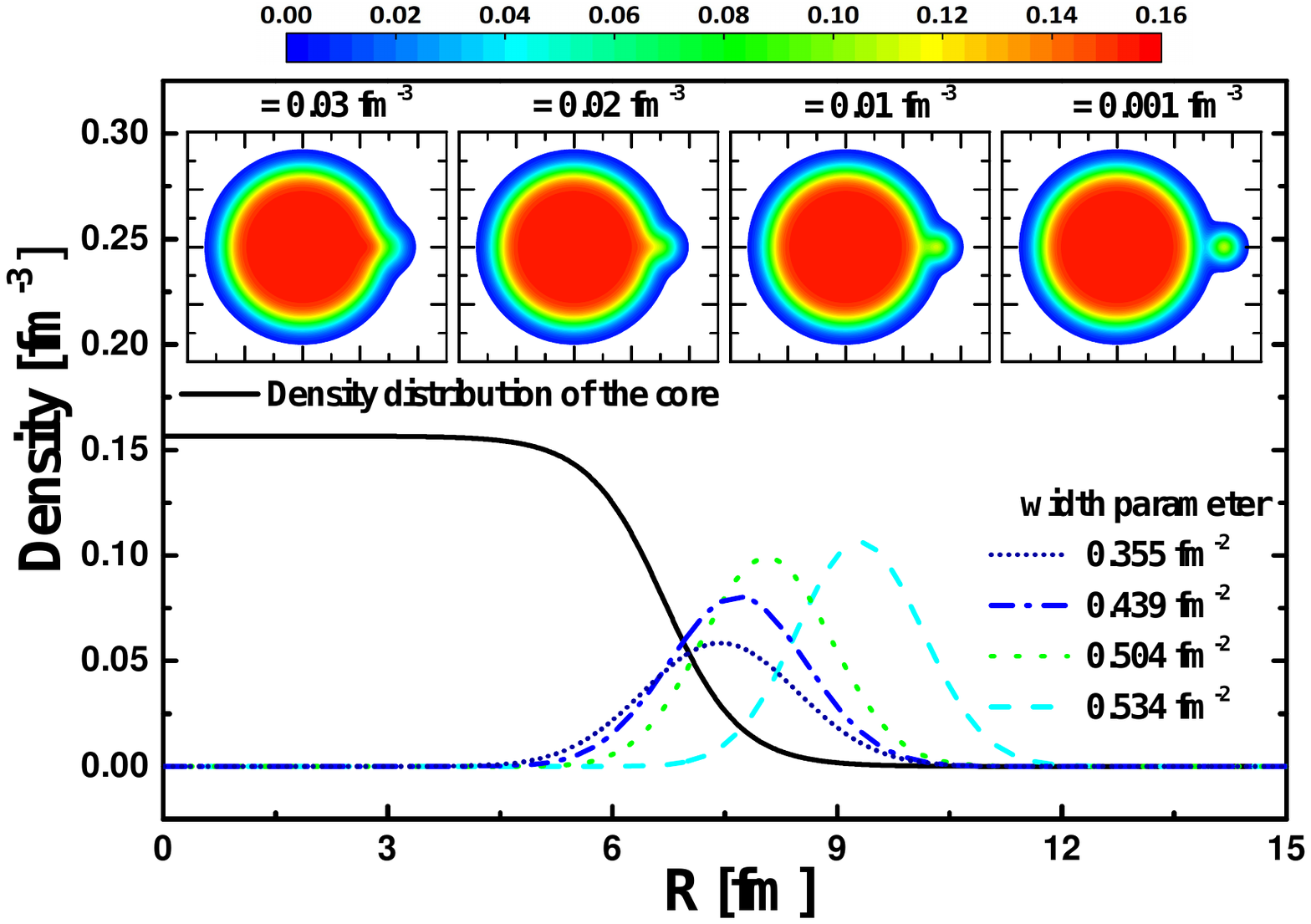}}
\end{minipage}
\caption{{Density evolution of an $\alpha$-cluster approaching the core nucleus. The width parameter $a(\rho)$ and density distribution of $\alpha$-cluster $\rho_{\alpha}(r,\rho)/4$ are shown for densities $\rho$=0.001 fm$^{-3}$, 0.01 fm$^{-3}$, 0.02 fm$^{-3}$, 0.03 fm$^{-3}$, respectively.}}
\label{density}
\end{figure}

{\it Density evolution of $\alpha$-cluster and formation probability in finite nuclei.}- The density evolution of an $\alpha$-cluster approaching the core nucleus is depicted in Fig.\ref{density}. The strong binding of $\alpha$-cluster is gradually reduced by the energy shift due to Pauli blocking after it feels the tail of the core density. As shown in Fig.\ref{density}, the variational parameter $a$ reflecting the size of $\alpha$-cluster is decreased from 0.534 to 0.355 when it merges with the continuum of single-particle states. Eventually the $\alpha$-cluster dissolves and its four nucleons go over into single-particle states with pair correlations in the open shells on top of the core. Before that, the $\alpha$-cluster remains a relatively compact entity with small extension even up to the critical density $\rho_c$ (see Fig.\ref{density}).

\begin{figure}[h]
\begin{minipage}{1\linewidth}
\centerline{\includegraphics[width=0.95\textwidth]{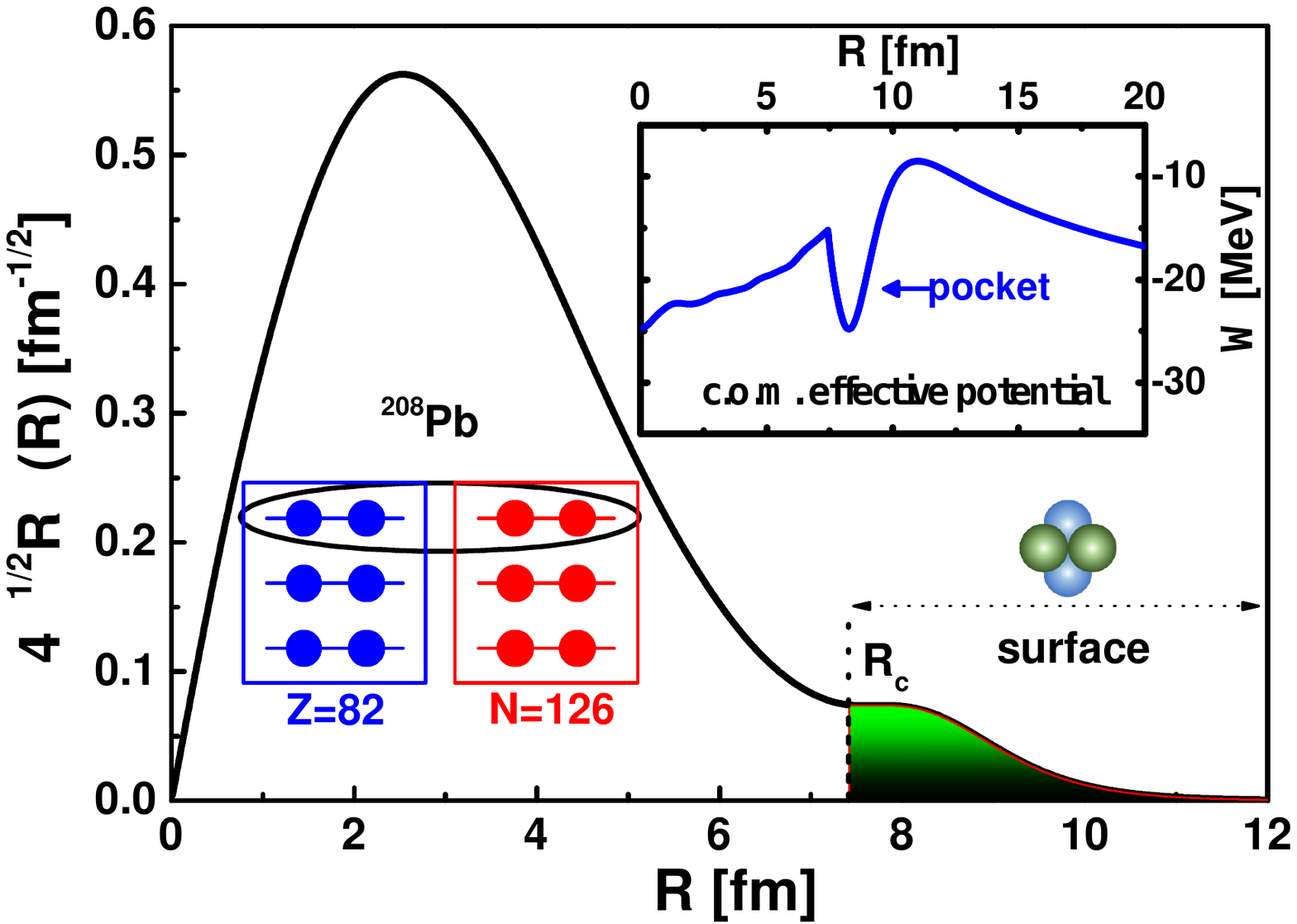}}
\end{minipage}
\caption{{The normalized c.o.m.\ wave function of four nucleons forming the $\alpha$-cluster. The wave function in the range of $0<R<R_C$ represents the c.o.m.\ wave function of four uncorrelated nucleons after dissolution. Only at the surface region with $R>R_C$, the $\alpha$-cluster appears and the c.o.m.\ wave function with $R>R_C$ corresponds to the formed $\alpha$-cluster, as marked by green color. The pocket region in the c.o.m.\ effective potential is denoted in the small panel.}}
\label{com potential}
\end{figure}

\begin{figure}[hbt]
\begin{minipage}{1\linewidth}
\centerline{\includegraphics[width=0.95\textwidth]{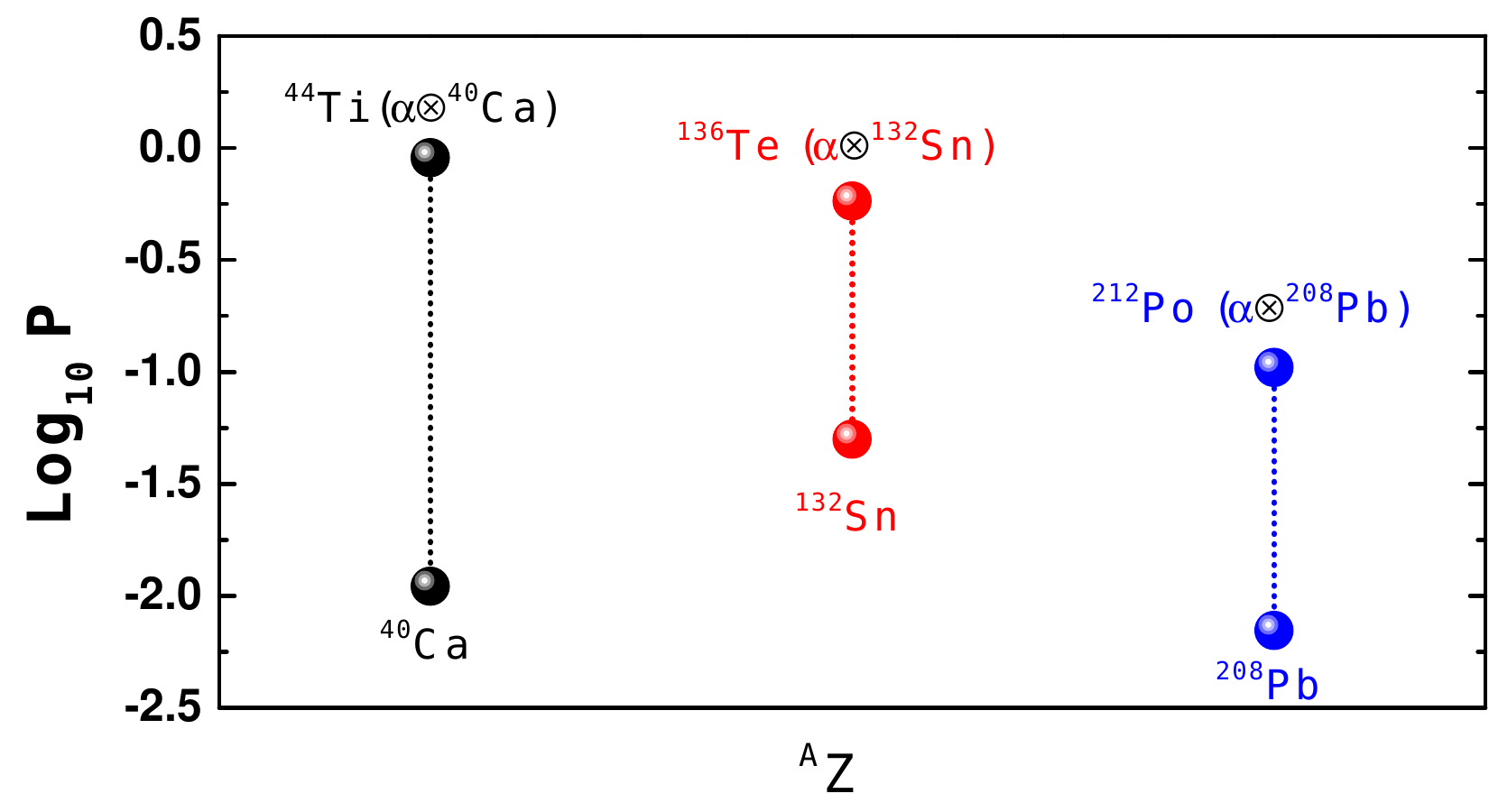}}
\end{minipage}
\caption{{Comparison of $\alpha$-cluster formation probabilities in doubly magic nuclei $^{40}$Ca, $^{132}$Sn, and $^{208}$Pb and in nuclei $^{44}$Ti, $^{136}$Te, and $^{212}$Po.}}
\label{magic}
\end{figure}

The c.o.m.\ motion of $\alpha$-cluster is introduced as a dynamical collective degree of freedom, which simplifies the treatment of correlated nuclear systems beyond the mean-field approximation. By separating the intrinsic motion from the c.o.m.\ motion, the c.o.m.\ wave function of $\alpha$-cluster follows the equation \cite{qwfa2020,qwfa2021},
\begin{widetext}
\begin{equation}
    \begin{aligned}
        &-\frac{\hbar^2}{2 Am}\nabla_R^2\Psi^{\text{com}}(\textbf{R})-\frac{\hbar^2}{A m}\int ds_j \varphi^{intr,\ast}(\textbf{R},\textbf{s}_j)[\nabla_R \varphi^{\text{intr}}(\textbf{R},\textbf{s}_j)][\nabla_R \Psi^{\text{com}}(\textbf{R})]\\
        &-\frac{\hbar^2}{2 Am}\int ds_j \varphi^{\text{intr},\ast}(\textbf{R},\textbf{s}_j)[\nabla_R^2\varphi^{\text{intr}}(\textbf{R},\textbf{s}_j)]\Psi^{\text{com}}(\textbf{R})+\int dR' W(\textbf{R},\textbf{R}')\Psi^{\text{com}}(\textbf{R}')=E\Psi^{\text{com}}(\textbf{R}),
    \end{aligned}
\end{equation}
\end{widetext}
where the second and third terms are complex derivative terms and no investigations of such terms have been performed in previous researches. It can be strictly proved that the second term vanishes if the number of nucleons ($A=4$) embedded in medium is conserved. In contrast, the third term is nontrivial and is rather difficult to solve (9-fold integral). For the first time, we take this derivative term into account in QWFA and find that this term does affect the final $\alpha$-cluster formation probability. The fourth term is the effective potential describing the c.o.m.\ motion of $\alpha$-cluster under the influence of Pauli blocking with the surrounding medium. The inner c.o.m.\ effective potential $W{(R<R_c)}$ ($R_c$ is the critical radius corresponding to $\rho_c$) is constructed from the shell model wave functions of four nucleons forming the $\alpha$-cluster. Note that only states near the Fermi energy can form an $\alpha$-like cluster because these shell model states extend to the low-density regions. The inner effective potential $W(R<R_c)$ joins with the outer one $W(R>R_c)={W(R)}^{\text{ext}}+{W(R)}^{\text{intr}}$ at $R=R_c$. An important feature of $W(R)$ is that a pocket is formed on the surface region (see the small panel in Fig.\ref{com potential}), resulting from the competition between strong nuclear force attraction and repulsive Pauli blocking. The pocket plays an essential role in the formation of $\alpha$-cluster at the surface of core nucleus. As seen in Fig.\ref{com potential}, the normalized c.o.m.\ wave function shows a small peak around the pocket region. {This is in agreement with the microscopic calculations on $\alpha$-clustering in Refs.\cite{Delion10,Delion04,Delion13}.} By integrating the c.o.m.\ wave function from the critical radius $R_c$ to infinity, the $\alpha$-cluster formation probability can be microscopically obtained $P_{\alpha}=\int_{R>R_c} d^3\text{R}|\Psi^{\text{com}}(\text{R})|^2$ \cite{qwfa2020,qwfa2021}. {We go beyond the Thomas-Fermi approximation by taking the closed shell structure effects into account. The $\alpha$-cluster formation probability is expected to vary dramatically across the major shell closures. Indeed, it is found that the $\alpha$-clustering in doubly magic nuclei like $^{40}$Ca, $^{132}$Sn, and $^{208}$Pb is significantly hindered by shell effects (see Fig.\ref{magic}). The {\it realistic} $\alpha$-cluster formation probability in $^{208}$Pb is rather small $P_{\alpha}=9.3\times10^{-3}$. This is quite different to their neighbors $^{44}$Ti, $^{136}$Te, and $^{212}$Po where enhanced $\alpha$-cluster formation probabilities are found by using exactly the same QWFA formulism.}

{\it Impact of $\alpha$-clustering on $R_{\rm skin}$ and $L$.}- A direct relationship between $L$ and the underlying single-nucleon potential $V_{n/p}(\rho,\delta,k)=V_0(\rho,k) \pm V_{\rm sym}(\rho,k)\delta$ is revealed by the HVH theorem \cite{symmetry,symmetry2014}. The advantage of this strict relationship is that it can be used to determine $L$ in a fully transparent way. At the saturation density $\rho_0$, $L$ can be reformulated by using the effective mass $m^*$,
\begin{eqnarray}
L=\frac{2}{3}t(k_F^0)+\frac{3}{2}V_{\rm sym}(\rho_0,k_F^0)+\frac{\partial V_{\rm sym}(\rho_0,k)}{\partial k}|_{k_F^0} k_F^0,
\label{Lformula}
\end{eqnarray}
where the first term $L(1)=\frac{\hbar^2 {k_F^0}^2}{3 m^*}$ denotes the contributions from kinetic energy and isoscalar potential \cite{symmetry}. For the nucleon effective mass, we adopt the value of $\frac{m^*}{m}$=0.70$\pm$0.05 widely used in the literature, see, {\it e.g.} Ref.\cite{Effmass}. {The isovector potential $V_{\rm sym}(\rho_0,k)$ can be deduced from the real part of global optical potentials, which is basically parameterized in the Woods-Saxon form, {\it i.e.} $V(r) = {-V_{0}[ 1\pm \kappa( \frac{N-Z}{A})]}/[{1+\textrm{exp}(\frac{r-R_0A^{1/3}}{a})}]$ (``$+$" for protons and ``$-$" for neutrons).} The second term $L(2)$ in Eq.(3) is determined by the product of the strength of the WS potential $V_0$ and the isovector parameter $\kappa$, {\it i.e.} $L(2)\!=\!\frac{3}{2}V_{\rm sym}(\rho_0,k_F^0)\!=\!\frac{3}{2}\kappa\cdot V_0$. The WS potential does not have explicit energy (or momentum)-dependence. From the global optical potential (GOP) constrained by nuclear reaction data \cite{symmetry}, the energy-dependence of the isovector potential is found to have a linear form $V_{\rm sym}(\rho_0,k)\!=\!22.75-0.21E(k)$ \cite{symmetry}. So the third term $L(3)$ in Eq.(3) is negative because of the decreasing isovector potential with increasing energy.

{We use the same ``QWFA" WS global optical potential to determine: 1) shell model states and densities in $^{208}$Pb and $^{48}$Ca, together with the Coulomb potential+$ls$ coupling 2) density slope of symmetry energy $L$ by using the HVH theorem. The ``QWFA"  parameterization is found to reproduce well the $\alpha$-cluster decay half-lives around doubly magic nuclei $^{208}$Pb and $^{100}$Sn \cite{qwfa2020,qwfa2021}.}
The neutron skin thickness $R_{\rm skin}=r_{n}^{\rm rms}-r_{p}^{\rm rms}$  is calculated directly from shell model density distributions. With explicit $\alpha$-cluster degree of freedom, the r.m.s.\ radius is given by $r^{\rm rms}=[\int r^2 (\rho^{\rm cluster} (r)+\rho^{\rm core}(r))d^3r]^{1/2}$ where $\rho^{\rm core}$ is the density distribution of protons or neutrons in the core. The density distribution of two neutrons or two protons forming the $\alpha$-cluster is
\begin{eqnarray}
\rho^{\rm cluster}(r)&=&2\int_{R<R_C}d^3\text{R}[\left|\Psi^{\text{com}}({\bf R})\right|^2\rho({\bf r})]
\\ \nonumber
&+&\frac{1}{2}\int_{R>R_C}d^3\text{R}[\left|\Psi^{\text{com}}({\bf R})\right|^2\rho_{\alpha}({\bf r}-{\bf R};{\bf R})],
\end{eqnarray}
where the $\alpha$-cluster formation at the surface of core nucleus ($R\ge R_c$) is taken into account in the second integral and the spatial extension of the formed $\alpha$-cluster is well described by $\rho_{\alpha}({\bf r}-{\bf R};{\bf R})$.
\begin{figure}[hbt]
\begin{minipage}{1\linewidth}
\centerline{\includegraphics[width=0.95\textwidth]{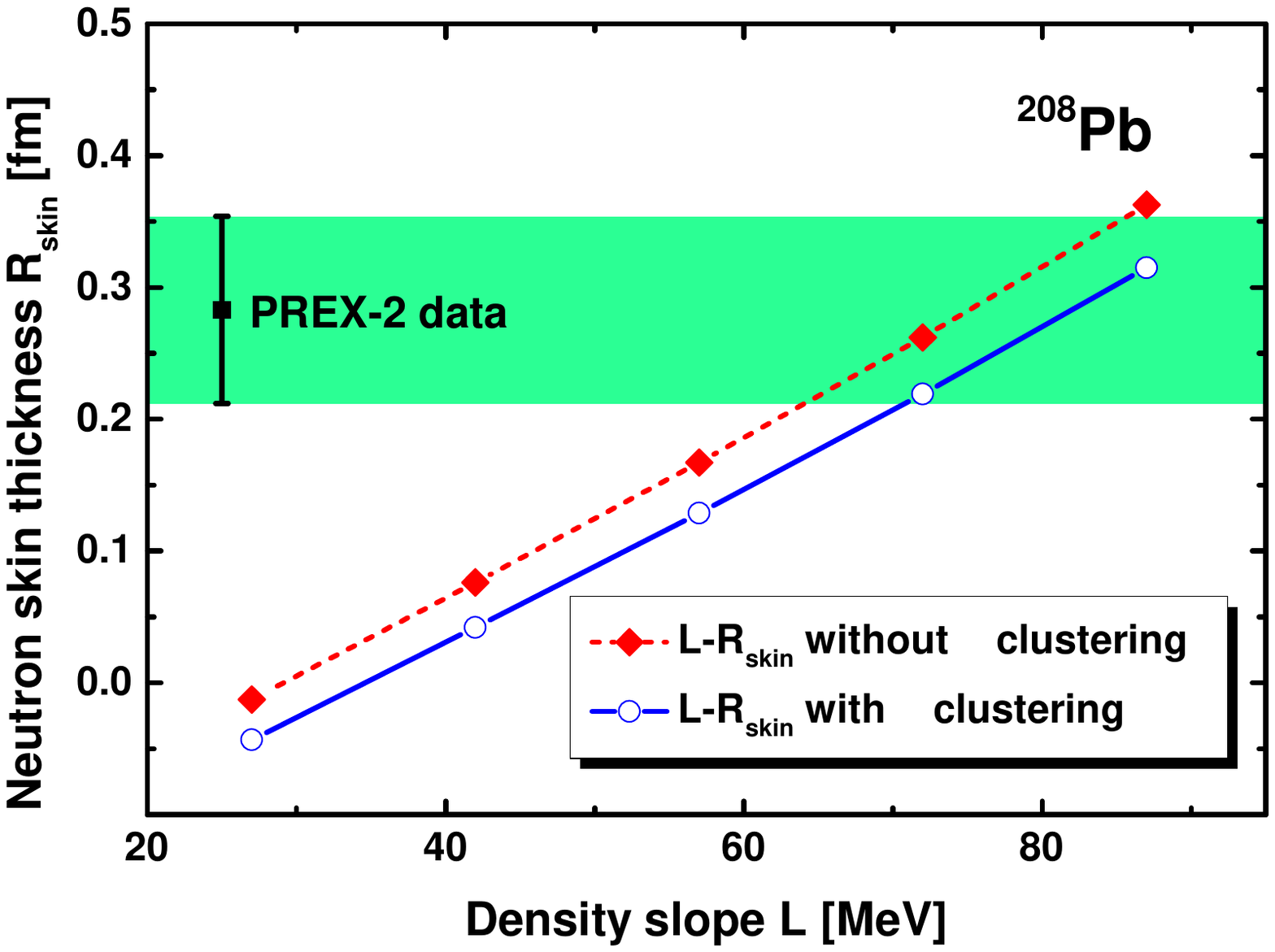}}
\end{minipage}
\caption{Correlation between $L$ and $R_{\rm skin}$ with and without $\alpha$-clustering in $^{208}$Pb. In the case of $\alpha$-clustering, a large amount of $\alpha$-cluster formation probability is assumed ($P_{\alpha}$=1).}
\label{Rskin}
\end{figure}
\begin{table}[htb]
\caption{The extracted density slope parameter $L$ by considering $\alpha$-clustering at the surface of $^{208}$Pb (PREX-2) and $^{48}$Ca (CREX).}
\begin{tabular}{|c|c|c|c|c|}
\hline
\multirow{2}{*}{Nuclei}     & \multirow{2}{*}{$R_{\rm skin}$ [fm]} & $L$ [MeV]           & \multirow{2}{*}{$P_{\alpha}$}             & $L$ [MeV]\\
           &                    & no $\alpha$-cluster &                        & with $\alpha$-cluster\\
\hline
$^{208}$Pb & 0.283$\pm$0.071 & 75.2$_{-24.5}^{+24.3}$  & 9.3$\times10^{-3}$        &75.3$_{-24.6}^{+24.3}$\\
\hline
$^{48}$Ca  & 0.121$\pm$0.050 & 13.2$_{-24.9}^{+25.4}$   & 7.3$\times10^{-2}$         &15.0$_{-25.0}^{+25.6}$\\
  {*}      & {0.071 (lower)}&  {1.7 }              &        & {3.4}\\
  {*}      & {0.171 (upper)}&  {24.8}             &       & {26.8}\\
\hline
\end{tabular}
{*The correction of $L$ due to $\alpha$-clustering  for the lower and upper limits of $R_{\rm skin}$ is 100\% and 8\%, respectively. }
\end{table}
We assume that the neutron skin thicknesses given by PREX-2 and CREX are all {\it measured} quantities. Fig.\ref{Rskin} shows the correlation between $L$ and $R_{\rm skin}$ with and without $\alpha$-clustering for $^{208}$Pb. $L$ increases with the increasing $R_{\rm skin}$ and the $L$-$R_{\rm skin}$ correlation is almost linear. We have checked the $L$-$R_{\rm skin}$ correlation by using different WS parameterizations \cite{SWV} and found that this behavior is general.
As shown in Fig.\ref{Rskin}, the $L$-$R_{\rm skin}$ correlation is modified significantly by assuming a large amount of formation probability ($P_{\alpha}$=1). However, the {\it realistic} $\alpha$-cluster formation probability in $^{208}$Pb is quite small, and thus its influence on $L$ is negligible. By considering the {\it realistic} $\alpha$-cluster formation probability in $^{208}$Pb, the $L$ value extracted from the PREX-2 data is 75.3 MeV (see Table I). The uncertainties of all terms contributing to this $L$ value are considered. The uncertainty in $L(1)$ is due to the effective mass $m^*$. With the $m^*/m$ = 0.70$\pm$0.05 we adopted, an error bar of $+$2.8/$-$2.4 MeV is obtained. Since the $R_{\rm skin}$ data of PREX-2 has large error bar, it is no surprise that there is large error bar associated with $L(2)$ term. The error bar associated with $L(3)$ term is obtained by the world data on nucleon-nucleus scatterings, (p, n) charge exchange reactions and single-particle energies of bound states \cite{symmetry}. Put all together the error bar of $L$ is approximately 24 MeV. In contrast to $^{208}$Pb, the {\it realistic} $\alpha$-cluster formation probability $P_{\alpha}$ in $^{48}$Ca is found to be 7.3$\times10^{-2}$, which is much larger than that in $^{208}$Pb. Thus the impact of $\alpha$-clustering in $^{48}$Ca on $L$ cannot be ignored. The extracted $L$ value with error bar from CREX data is 15.0$_{-25.0}^{+25.6}$ MeV and the correction due to $\alpha$-clustering is of the order of 14\%. {This correction progressively increases with the $\alpha$-cluster formation probability $P_{\alpha}$, which could be close to unity if the contributing shell model orbits are rather similar, especially for self-conjugate nuclei such as $^{44}$Ti in Fig.\ref{magic}.}

{\it Conclusion.}- $\alpha$-clustering survives at the surface of heavy nuclei, which is relevant to the neutron skin thickness of heavy nuclei. The latter is a precise tool in constraining the density slope of nuclear symmetry energy. The impact of $\alpha$-clustering on the neutron skin thickness depends closely on the amount of the formation probability. We emphasize that the approach presented here to calculate formation probability is based on a first-principle approach to nuclear many-body systems. A proper treatment of derivative terms of intrinsic wave function has been performed and the spatial extension of the $\alpha$-cluster has been considered to better account the correlation between $L$ and $R_{\rm skin}$. Present analysis shows that the $L$ values deduced from  PREX-2 and CREX experiments are not consistent with each other, even with $\alpha$-clustering effect included. We expect that a better account of model-dependence in extracting $R_{\rm skin}$ from parity-violating asymmetry $A_{PV}$ will further improve the estimation of $L$. Moreover, state-of-art approaches can be applied to describe shell model states and nuclear densities and the Gaussian ansatz used in the variational calculations can be improved. Exact solution of the four-nucleon correlation near the critical density by including both self-energy corrections and Pauli blocking should be tackled in future.

{\it Acknowledgments.}- Discussions with G. R\"opke, Z. Ren, Y. Funaki, H. Horiuchi,
A. Tohsaki, T. Yamada, B. Zhou, L. W. Chen and C. D. Roberts are gratefully acknowledged. This work is supported by the National Natural Science Foundation of China (Grant No. 11822503).

\end{document}